\documentclass[aps,pra,twocolumn,showpacs,superscriptaddress,groupedaddress]{revtex4}  
\usepackage{graphicx}  
\usepackage{dcolumn}   
\usepackage{bm}        
\usepackage{amssymb}   
\usepackage{amsmath}
\usepackage{dsfont}
\usepackage{cancel}
\usepackage{empheq}
\usepackage{graphicx,color}
\usepackage{array}
\usepackage{tikz,pgf}
\def\ket#1{| \,#1\, \rangle}
\def\bra#1{\langle \,#1\, |}

\begin{document}



\title{Likelihood Theory in a Quantum World: tests with Quantum coins
and computers}
\author{Arpita Maitra$^{}$}
\address{${}$ Indian Institute of Technology, Kharagpur}
\author{Joseph Samuel$^{}$}
\author{Supurna Sinha$^{}$}
\address{${}$Raman Research Institute, Bangalore 560080, India.
}



\date{\today}

\begin{abstract}
By repeated trials, one can determine the fairness of a classical coin
with a confidence which grows with the number of trials. 
A quantum coin can be in a superposition of heads and tails and its state is most generally
a density matrix. Given a string of qubits representing a series of trials, 
one can measure them individually and determine the state with a certain confidence. We show that there is
an improved strategy which measures the qubits after entangling them, which leads to a greater
confidence. This strategy is demonstrated on the simulation facility of 
IBM quantum computers. 
\end{abstract}

\pacs{03.67.-a, 03.67.Lx, 03.65.Ta, 03.65.Ud}
\maketitle

{\it{Introduction:}}
When testing a theory against experimental data, 
the Bayesian approach gives us a rational way
of revising our theoretical expectations in the light of 
new data. To take the simple and familiar example of coin tossing,
let us start with the belief that our coin is fair. If we then toss the
coin ten times and turn up nine heads, our belief in fairness will be 
shaken but not destroyed: it is still possible that the nine heads
were generated by chance. If the run of heads continues, we would
be hard pressed to cling to our belief in the fairness of the coin.
Revising our beliefs in the light of new data is an essential component of
the scientific method, a point which was strikingly brought out in 
the Monty Hall problem, which occupied the community in the 1990's.
Bayesian theory is routinely used in testing the efficacy of drugs, 
where one starts with the ``null hypothesis'' that the 
drug being tested is no more effective than a placebo. If the 
drug fares better than a placebo over a sufficiently large number 
of trials, we are more inclined to  believe in its efficacy.

Let us now return to coin tossing, which captures the idea of Bayesian
inference in its simplest form.
The intuitive idea is quantitatively captured in likelihood theory. 
Our initial belief (prior) is that the probability of 
heads is $q_{H}$ and tails $q_{T}$. If we see a particular 
string ${\cal S}_N=\{HTHHHHTTHTTT.....\}$, in $N$ coin tosses, the likelihood of such
a string emerging from the distribution $P_B=\{q_H,q_T\}$ 
is given by (see below
for a derivation)
\begin{equation}
L({\cal S}_N|P_B)=\frac{1}{\sqrt{2\pi Np_H p_T}} \exp{-\{ND_{KL}(P_A\|P_B)\}}
\label{firstequation}
\end{equation}
where $P_A=\{p_H,p_T\}=\{N_H/N,N_T/N\}$ is the observed frequency distribution
in a string of $N$ tosses. $D_{KL}(P_A\|P_B)$ is the relative entropy or the 
Kullback-Leibler divergence between the distributions $P_B$ and $P_A$.

When we pass from classical coins to quantum coins, a new possibility emerges: the coins may 
be in a superposition of heads $\ket{H}$ and tails $\ket{T}$. Even more generally, the state of the system need not
even be pure but a density matrix.
The probability distributions $P_B, P_A$
describing classical coins 
are replaced by density matrices $\rho_B$ and $\rho_A$. 
One of the central problems in quantum information is quantum state discrimination. Given two quantum states, 
how easily can we tell them apart? This question has led to the issue of distinguishability measures
on the space of quantum states \cite{shunichi,statest, anthony, osaki, barnett}. More recently in \cite{entropy} 
it has been noticed that quantum entanglement leads to a significant improvement in distinguishing two quantum 
states.
In this Letter, we go beyond \cite{entropy} by actually implementing our theoretical model on a state of the art quantum computer, 
the IBM quantum computer and checking our theoretical ideas 
against quantum simulations and experiments.

Our goal in this Letter is to describe an experiment which demonstrates the use of entanglement as a resource 
in state discrimination.  We have successfully implemented our ideas on the ibm quantum simulator. In experimental
runs, it appears that the quantum advantage we seek to demonstrate is swamped by noise.
The experiment may well be within the reach of more sophisticated quantum computers which are currently in use
and so provides a short term goal possibly within reach.

The outline of this Letter is as follows. 
 We first review classical likelihood theory 
to set the background for our work.  
We then describe how likelihood theory is modified in a quantum world 
and show how these abstract ideas can be
translated into reality using existing quantum computers.
We discuss the simplest direct measurement strategy where one 
measures qubits one by one.
We then show that there exist improved strategies
which exploit quantum entanglement in an essential way.
We translate our abstract ideas into 
quantum scores, combinations of quantum logic gates, which can be 
run on quantum computers. 
Finally, we present the results of running 
our programs on existing IBM quantum computers
and  end with some concluding remarks.

{\it{Review of Classical Likelihood Theory:}}
Let us consider a biased coin for which the probability of getting a head is 
$p_H= 1/3$ and that of getting a tail is $p_T= 2/3$. Suppose we incorrectly assume 
that the coin is fair and assign probabilities $q_H= 1/2$ 
and $q_T= 1/2$ for getting a head and a tail respectively. 
The question of interest is the number of trials needed to be able to distinguish
(at a given confidence level)
between our assumed probability distribution and the measured probability distribution. 
A popular measure for distinguishing between the expected distribution and the measured distribution is 
given by the relative entropy or the KL divergence (KLD) 
which is widely used in the context of distinguishing
classical probability distributions \cite{jon}.
Let us consider $N$ independent tosses of a coin leading to a string 
${\cal S}_N=\{HTHHTHTHHTTTTT......\}$. 
What is the probability that the string ${\cal S}$ is generated by the model distribution $P_B=\{q,1-q\}$?
The observed frequency distribution is $P_A=\{p,1-p\}$. 
If there are $N_H$ heads and $N_T$ tails in the  string ${\cal S}_N$ 
then the probability of getting such a string is  $\frac{N!}{N_H!N_T!}q^{N_H}(1-q)^{N_T}$ 
which we call the likelihood function $L({\cal S}_N|P_B)$. 
If we take the average of the logarithm of this likelihood function and use Stirling's 
approximation for large $N$ we get the following expression:
\begin{equation}
\frac{1}{N}\log{L(N|{P_B)}}=-D_{KL}(P_A\|P_B)+\frac{1}{N}\log{\frac{1}{\sqrt{2\pi Np(1-p)}}},
\label{likelihood}
\end{equation}
where $p=\frac{N_H}{N}$ and $D_{KL}(P_A\|P_B)=p\log{\frac{p}{q}}+(1-p)\log{\frac{1-p}{1-q}}$.
The second term in (\ref{likelihood}) is due to the sub-leading term $\frac{1}{2}\log{2\pi N}$ of Stirling's approximation. 
If $D_{KL}(P_A\|P_B)\neq 0$ then the likelihood of the string $S$ being produced by the $P_B$ distribution 
decreases exponentially with $N$. This results in eq.(\ref{firstequation}).
Thus $D_{KL}(P_A\|P_B)$ gives us the divergence of the measured distribution from the model distribution.
The KL divergence is positive and vanishes if and only if the two distributions $P_A$ and $P_B$ are equal. 

In this limit,
we find that  
the exponential divergence gives way to a power law divergence, as shown by the subleading term in (\ref{likelihood}).
The arguments above generalize appropriately to an arbitrary number of outcomes (instead of two) and also 
to continuous random variables.

{\it{Quantum Likelihood Theory:}}
In passing from classical likelihood theory described above to the quantum one,
there are two additional features to be considered. First, 
as mentioned earlier, because of the superposition principle,
a quantum coin would in general be described by a 
density matrix rather than a probability distribution. Second, in measuring 
a quantum system, one has an additional freedom: the choice of 
measurement basis. It is evident that the choice of measurement basis can  
affect our discriminating power between competing quantum states. 
We would be well advised to choose the basis wisely, so as to maximise our
discriminating power between two given density matrices.

Consider two mixed density matrices $\rho_A$ and $\rho_B$. 
Suppose that the (prior) quantum distribution is described by  
$\rho_B$. For a choice $\{m_i\}$ ($i=1...n$) of orthonormal basis, 
the probable outcomes have the probabilities $q_i=\bra{m_i}\rho_B\ket{m_i}$. 
Similarly, if the density matrix is $\rho_A$,
we have the probabilities  $p_i=\bra{m_i}\rho_A\ket{m_i}$. Both $\{q_i\}$ and $\{p_i\}$ are normalised
probability distributions. These can now be regarded as classical probability
distributions and we can compute their
relative entropy, defined as
$D_{KL}(m) = \sum_{i=1..n} p_i Log(p_i/q_i)$.
The argument $m$ on the LHS indicates that the relative entropy {\it depends} on the measurement basis $m$, since the $p$s and $q$s do.
Let us now choose $m=m^*$ optimally so as to maximise $D_{KL}(m)$ 
which is a measure of our discriminating power.  The basis optimised
value of the Kulback-Leibler divergence is denoted by $S(A||B)= D_{KL}(m^*)$.
For example, we could be detecting a weak signal (as in quantum metrology) using a quantum detector, which
would remain in the state $\rho_B$ in the absence of an incoming signal, but get excited to state $\rho_A$ when the signal is
present. If the state is $\rho_A$, our confidence $C$ that the state 
is in fact $\rho_B$ will decrease at the rate
$$C=\frac{1}{\sqrt{2\pi Np(1-p)}} \exp{-\{N S(A||B)\}}.$$
where $N$ is the number of measured systems.

{\it{Tossing Quantum Coins:}}
The abstract theory of the last section can be converted into concrete experiments using 
the recent progress in constructing quantum computers in the laboratory. Choosing the optimal basis is 
effected by a Unitary transformation, which can be implemented using quantum gates.
The quantum circuit that we implement on the quantum computer can
be divided into three parts. The first part is state preparation,
the second is Unitary transformation and the third is measurement.
We explore two strategies, which we call the direct and the entangling strategy.
The two strategies differ only in the second part. In the direct
strategy, we measure the qubits one at a time, taking care to 
choose the measurement basis so as to maximise our ability to distinguish
the two states. In  the entangling strategy, we perform an entangling unitary 
transformation on a pair of qubits before 
performing the measurement.

{\it State Preparation:} 
Without loss of generality, we can assume the state
$\rho_A$ to be along the $z$ direction of the Bloch sphere 
and $\rho_B$ to lie in the $x-z$ plane of the Bloch sphere, 
making an angle $\delta$ with the $z$ axis.
To prepare the initial quantum state $\rho_A$, which is a density matrix, 
we begin with two qubits (say a,b) both in the state $\ket{0}$:
$\ket{0}_a\otimes\ket{0}_b$. 
We then perform a local unitary rotation through an angle $\beta$ 
in the $x-z$ plane on one of them (b) and 
arrive at $\ket{0}_a\otimes(\cos{\beta/2}\ket{0}_b+\sin{\beta/2}\ket{1}_b)$. We then 
apply a CNOT gate with b as the control and a as the target. 
The total system is now in an entangled state. If we ignore (trace over) qubit a, we get an impure density matrix on qubit b. 
$\rho_A=(\cos{\beta/2})^2\ket{0}\bra{0}+(\sin{\beta/2})^2\ket{1}\bra{1}$
This is our initial state $\rho_A$, whose purity depends on the entangling angle $\beta$. In runs we have used the value of
$\beta=0.2$ and $\delta=1.8$.

In practice (and for easier comparison with the entangling strategy) 
we repeat the arrangement with two more qubits with qubits a and b 
replaced respectively by d and c. We get  identical states in qubits b and c. 
The rest of the circuit performs an unitary transformation on the qubits b and c and 
then performs a measurement in the computational basis.

{\it{Direct strategy:}}
In the direct measurement strategy we perform measurements on individual qubits.  One can show \cite{supplementary}
that the optimal basis consists of two antipodal points in the $x-z$ plane of the Bloch sphere. This determines a direction
in the $x-z$ plane which is characterised by an angle $\varphi$ in the range $0-\pi$.
We optimise over $\varphi$ by choosing $\varphi$, the measurement basis so that $D_{KL}(\varphi)$ defined in the last section, 
is as large as possible. 
Then $$S_{rel}=D_{KL}(\varphi^{*})$$ is the maximum value where $\varphi^{*}$ is the value of $\varphi$ corresponding to the maximum 
value of $D_{KL}$. This gives us the optimal one-qubit strategy 
for distinguishability of $\rho_A$ and $\rho_B$. This is 
our direct measurement strategy.   
The optimal unitary transformation for the direct strategy is easy to find numerically. The unitary transformations of interest are in fact 
orthogonal \cite{supplementary}, and belong to $SO(2)$ (because the states are assumed to be in the $x-z$ plane) and characterised by a single angle $\varphi$. We numerically evaluate
the relative entropy as a function of $\varphi$ and find the maximum $\varphi^*$. We use this unitary transformation $U_2$ to get the optimised
measurement basis, which is implemented as a local transformation on 
qubits b and c. The logic circuit which implements the direct strategy
on a quantum computer is displayed in Fig. 1.

{\it{Entangling strategy:}}
The entangling strategy  
for discriminating qubits gives us an advantage over 
the Direct Measurement Strategy, in 
quantum state discrimination. 
Consider $2N$ idependent and identically distributed qubits. 
Let us group these $2N$ qubits into $N$ pairs. 
We now perfom measurements on each pair, choosing our measurement 
basis to give us optimal results for state distinguishability. This is effected by performing an unitary 
transformation $U_4$ {\it in the two qubit Hilbert space} followed by measurement in the computational basis. By grouping the qubits in pairs,
we have increased the freedom in the choice of measurement basis. For, we can choose entangled bases as well 
as separable ones. This greater freedom means that we can improve on the direct strategy. 
Maximising $S_{rel}$ over {\it all} two qubit bases, is clearly an improvement over maximising over 
separable ones. 

We maximise the relative entropy  in the space of two qubit bases, by doing a numerical search. Computer experiments
reveal that one does indeed gain by using the entangling strategy.
Our initial aim was to show that this entangling strategy 
indeed gives us an improvement over the Direct Measurement Strategy which we 
have described in the last section, by explicitly implementing it on a quantum computer.   

For the entangled strategy, we find it advantageous to simplify our search 
by searching within $SO(4)$. We perform a random walk in the 
$SO(4)$ space to find the optimal $U_4$. This yields a relative entropy per qubit which is higher than
that obtained by the direct strategy. It is known from the theory of 
Makhlin transforms \cite{makhlin,shende} how to decompose any $U_4$ into
local unitary gates and CNOT gates. It turns out that the unitary can be implemented (up to a phase) by using just two CNOT gates
and six unitary gates. Our mathematica programs compute the required unitaries and produce a qasm program that can be implemented on 
the ibm machines. The quantum logic circuit implementing the entangling strategy is shown in Fig. 2.

\begin{figure}[h] 
\begin{center}
\includegraphics[width=.56\textwidth,trim = 1cm 10cm 1cm 1cm,clip]{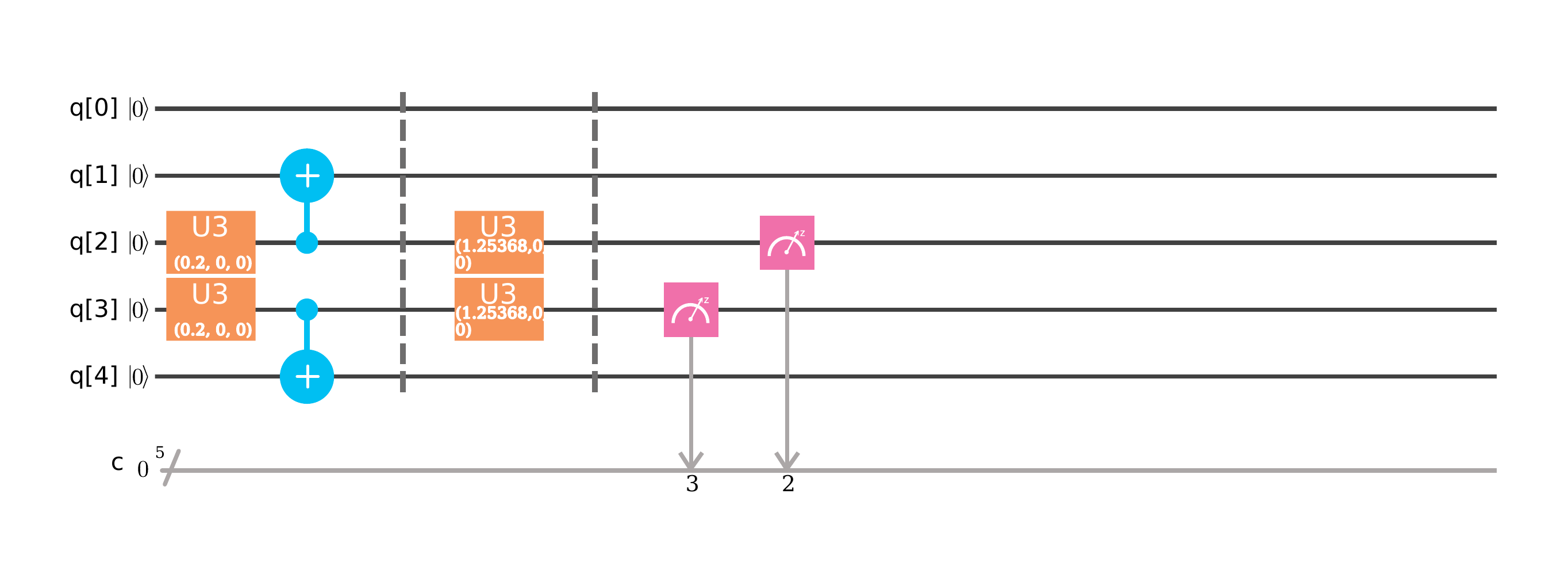}
\caption{The quantum logic circuit for the direct strategy. 
The unitary transformation to the optimal separable basis is between the two black dashed vertical lines.
The preparation stage is the part to the left of these lines  and to the right of these lines, the pink 
blocks with dials show the measurement stage.}
\label{1}
\end{center}
\end{figure}

\begin{figure}[h] 
\begin{center}
\includegraphics[width=.56\textwidth,trim = 1cm 10cm 1cm 1cm,clip]{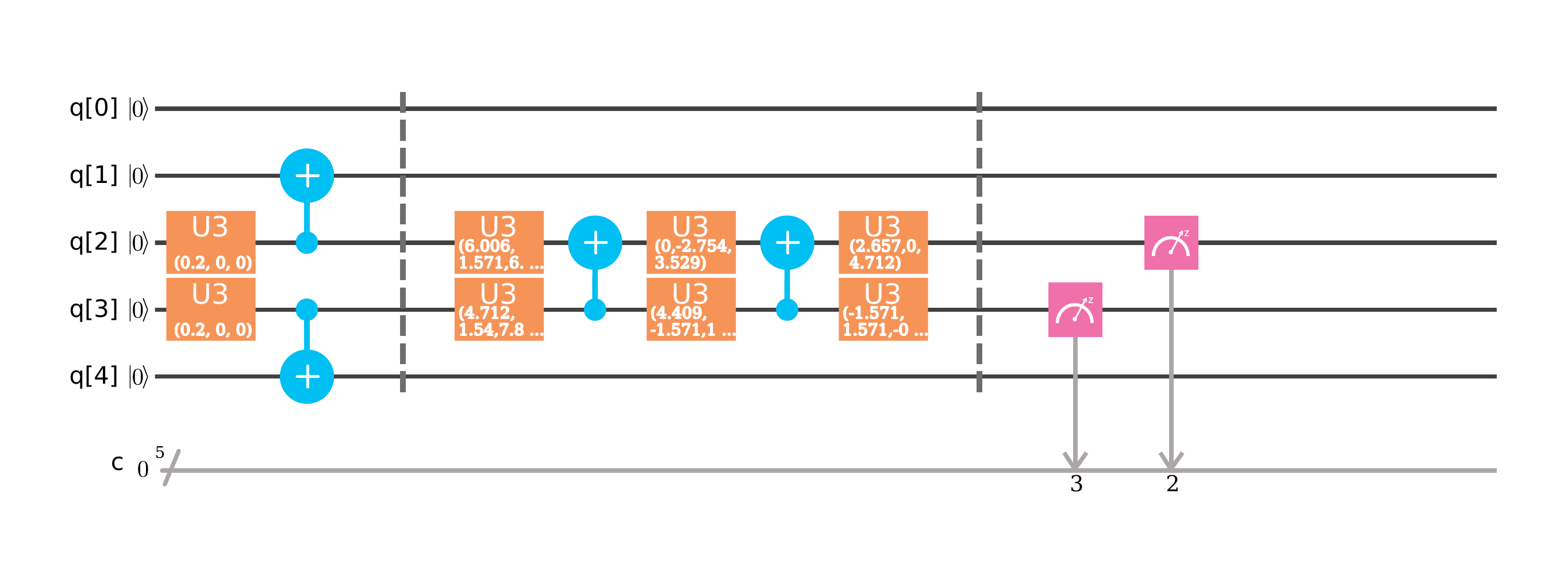}
\caption{The quantum logic circuit for the entangling strategy. 
The unitary transformation to the optimal entangling basis is between the two black dashed vertical lines.
This entangling unitary transformation is
realised (upto an overall phase) as a string of unitary operations and two CNOT gates.
The preparation stage is the part to the left of these  lines and to the right of these lines,
the pink blocks with dials show the measurement stage.}
\label{2}
\end{center}
\end{figure}
\begin{figure}[h] 
\begin{center}
\includegraphics[width=0.4\textwidth]{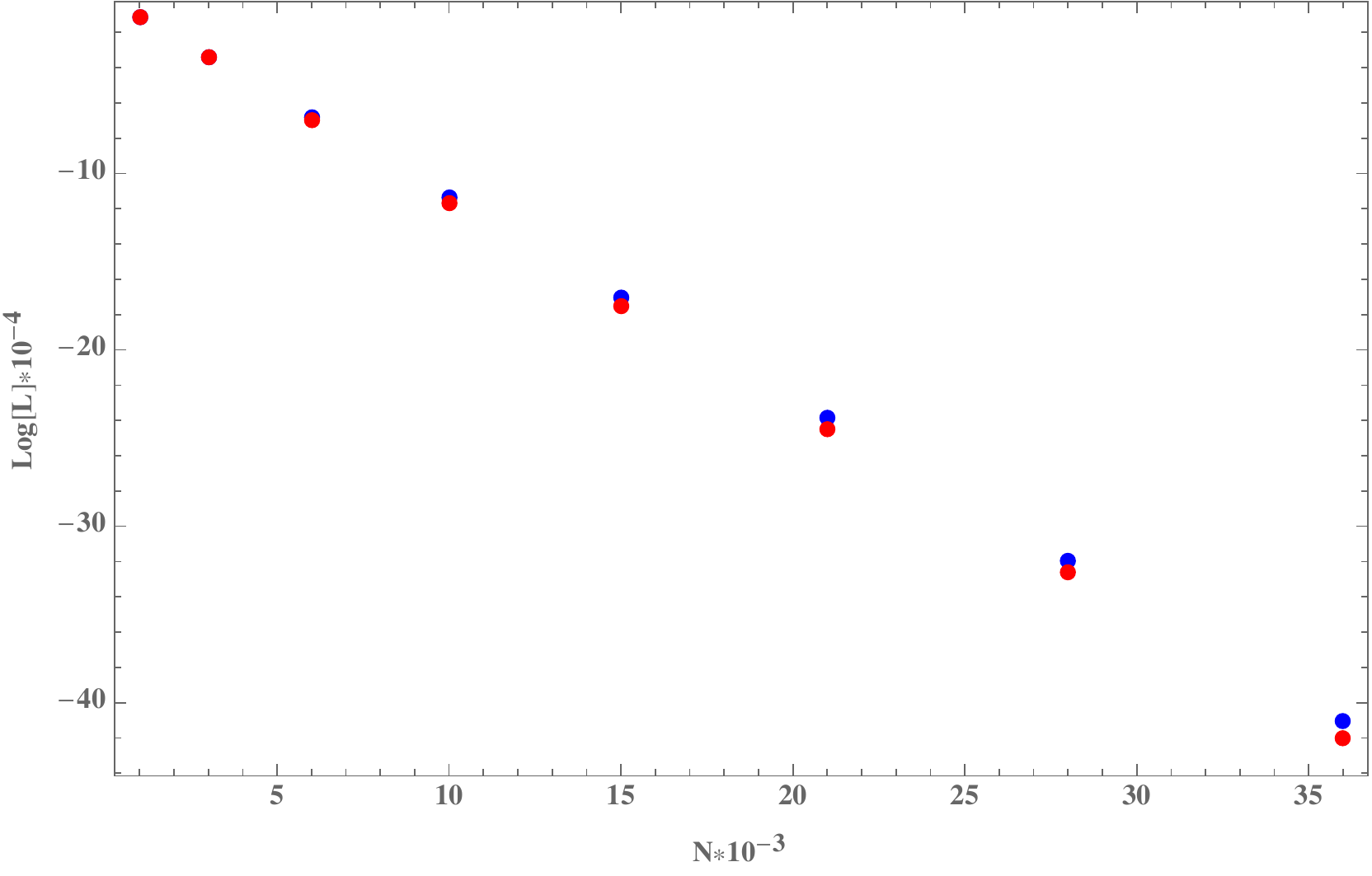}
\caption{Figure shows the log of the likelihood as a function of the number
or runs. The red dots represent the result of using the entanglement strategy
and the blue ones use the direct strategy.  The red dots show a steeper 
negative slope. This shows that the likelihood of the state being $\rho_B$ 
decreases faster as the number of qubits measured increases.
}
\label{3}
\end{center}
\end{figure}

{\it {Results:}}
Let us summarize the results of the study. We have implemented our idea at three levels:

(a) a mathematica program which numerically computes and compares the relative entropy for the two computational strategies - the Direct and the Entangled,
(b) a simulation on the ibmqx4 computer which mimics the actual quantum experiment by incorporating some of the specific engineering details, and 
(c) an actual quantum experiment on the ibmqx4 computer.
Below is a table which summarizes our results. 
The relative entropies corresponding to the direct and entangling strategies $S_{\mathrm{reldir}}$
and $S_{\mathrm{relent}}$ and their difference $S_{\mathrm{reldiff}}=S_{\mathrm{relent}}-S_{\mathrm{reldir}}$ 
are displayed 
corresponding to the mathematica program (Theory), ibm simulator (Sim) and ibm experiment (Expt) in the table.



\noindent

\begin{center}
\begin{tabular}{ | m{5em} | m{2cm}| m{2cm} | m{2cm}| } 
\hline
& Theory& Sim &Expt\\ 
\hline
Srel dir & 4.506& 4.519&4.043 \\ 
\hline
Srel ent &  4.723 &  4.696&1.660\\ 
\hline
Srel diff&  0.217&  0.177&-2.382\\ 
\hline
\end{tabular}
\end{center}

We notice that our mathematica programs are in close agreement with the results of the simulations on the ibmqx4 quantum computer. However, there is a considerable 
discrepancy between the results of the real quantum ibmqx4 experiments and the ibmqx4 simulations or the results of the mathematica programs. This discrepancy 
could  be due to limitations coming from gate infidelity and decoherence. Thus the noise from these two sources masks 
the quantum advantage effect that we expect from our theoretical studies 
\cite{entropy}.

We graphically capture the main result of our study in Fig.3 where we plot the Log[Likelihood] versus the number of qubits which clearly demonstrates the superiority
of the Entangled Strategy over the Direct Strategy. Thus there is a clear quantum advantage in choosing the Entangled Strategy instead of the Direct one.

{\it{Conclusion:}}
In this Letter, we described how Likelihood theory works in
the quantum regime and show how these ideas can be implemented on a quantum
computer. The main message is that entanglement can be used as a resource
to improve our power to discriminate between quantum states of qubits. Entanglement is crucially
used here, since it is our ability to use non separable bases that gives us the quantum advantage.

The scheme we propose has been implemented
on the ibmqx4 computer. We have performed both simulations and real experiments
on this computer. We notice that our theoretical expectation based on Mathematica
programs agrees very well with results of the simulations. However, decoherence and 
limitations in gate fidelity mask the effect of quantum advantage stemming from 
entanglement that we notice in our Mathematica programs and simulations on ibmqx4.
We have also tested our ideas on the Melbourne quantum computer IBM Q16, 
which has provided some marginal improvement in the experimental results.

Our study therefore has two significant aspects to it. It has provided a way to 
check theoretical expectations of quantum advantage from entanglement in the context 
of Quantum Likelihood Theory. In addition it has brought out the limitations of the 
present day ibmqx4 computer so far as experiments go. This would motivate researchers 
to improve the quality of the experiments by improving gate fidelity and by reducing 
noise due to decoherence.

{\it{Acknowledgement}}
 We acknowledge the use of the ibmqx4 quantum computer and 
the IBM Q16 computer in Melbourne.

\begin{thebibliography}{10}
\expandafter\ifx\csname natexlab\endcsname\relax\def\natexlab#1{#1}\fi
\expandafter\ifx\csname bibnamefont\endcsname\relax
  \def\bibnamefont#1{#1}\fi
\expandafter\ifx\csname bibfnamefont\endcsname\relax
  \def\bibfnamefont#1{#1}\fi
\expandafter\ifx\csname url\endcsname\relax
  \def\url#1{\texttt{#1}}\fi
\expandafter\ifx\csname urlprefix\endcsname\relax\def\urlprefix{URL }\fi
\providecommand*{\bibinfo}[2]{#2}
\providecommand*{\eprint}[1]{#1}
\providecommand*{\url}[1]{#1}
\begingroup\makeatletter
 \@temptokena{%
  \expandafter\ifx\csname citenamefont\endcsname\relax
   \DeclareRobustCommand\citenamefont{\@firstofone}%
   \global\let\citenamefont\citenamefont
   \global\expandafter\let\csname citenamefont \expandafter\endcsname\csname
  citenamefont \endcsname
  \fi
 }\if@filesw\immediate\write\@auxout{\the\@temptokena}\fi
\expandafter\endgroup\the\@temptokena

\bibitem[{\citenamefont{Amari}(2016)}]{shunichi}
\bibinfo{author}{\bibfnamefont{S.}~\bibnamefont{Amari}},
  \emph{\bibinfo{title}{Information Geometry and Its Applications}}, Applied
  Mathematical Sciences (\bibinfo{publisher}{Springer Japan},
  \bibinfo{year}{2016}), ISBN \bibinfo{isbn}{9784431559788},
  \urlprefix\url{https://books.google.co.in/books?id=UkSFCwAAQBAJ}.

\bibitem[{\citenamefont{Teo}(2015)}]{statest}
\bibinfo{author}{\bibfnamefont{Y.~S.} \bibnamefont{Teo}},
  \emph{\bibinfo{title}{Introduction to Quantum-State Estimation}}
  (\bibinfo{publisher}{World Scientific}, \bibinfo{year}{2015}), ISBN
  \bibinfo{isbn}{9789814678834}.

\bibitem[{\citenamefont{Chefles}(2000)}]{anthony}
\bibinfo{author}{\bibfnamefont{A.}~\bibnamefont{Chefles}},
  \bibinfo{journal}{Contemporary Physics}
  \textbf{\bibinfo{volume}{41}}(\bibinfo{number}{6}), \bibinfo{pages}{401}
  (\bibinfo{year}{2000}),
  \urlprefix\url{http://dx.doi.org/10.1080/00107510010002599},
  \eprint{http://dx.doi.org/10.1080/00107510010002599}.

\bibitem[{\citenamefont{Osaki} \emph{et~al.}(1996)\citenamefont{Osaki, Ban, and
  Hirota}}]{osaki}
\bibinfo{author}{\bibfnamefont{M.}~\bibnamefont{Osaki}},
  \bibinfo{author}{\bibfnamefont{M.}~\bibnamefont{Ban}}, \bibnamefont{and}
  \bibinfo{author}{\bibfnamefont{O.}~\bibnamefont{Hirota}},
  \bibinfo{journal}{Phys. Rev. A} \textbf{\bibinfo{volume}{54}},
  \bibinfo{pages}{1691} (\bibinfo{year}{1996}),
  \urlprefix\url{http://link.aps.org/doi/10.1103/PhysRevA.54.1691}.

\bibitem[{\citenamefont{Barnett and Croke}(2009)}]{barnett}
\bibinfo{author}{\bibfnamefont{S.~M.} \bibnamefont{Barnett}} \bibnamefont{and}
  \bibinfo{author}{\bibfnamefont{S.}~\bibnamefont{Croke}},
  \bibinfo{journal}{Adv. Opt. Photon.}
  \textbf{\bibinfo{volume}{1}}(\bibinfo{number}{2}), \bibinfo{pages}{238}
  (\bibinfo{year}{2009}),
  \urlprefix\url{http://aop.osa.org/abstract.cfm?URI=aop-1-2-238}.

\bibitem[{\citenamefont{Shivam} \emph{et~al.}(2018)\citenamefont{Shivam, Reddy,
  Samuel, and Sinha}}]{entropy}
\bibinfo{author}{\bibfnamefont{K.}~\bibnamefont{Shivam}},
  \bibinfo{author}{\bibfnamefont{A.}~\bibnamefont{Reddy}},
  \bibinfo{author}{\bibfnamefont{J.}~\bibnamefont{Samuel}}, \bibnamefont{and}
  \bibinfo{author}{\bibfnamefont{S.}~\bibnamefont{Sinha}},
  \bibinfo{journal}{International Journal Of Quantum Information}
  \textbf{\bibinfo{volume}{16}}, \bibinfo{pages}{1850032}
  (\bibinfo{year}{2018}).

\bibitem[{\citenamefont{Shlens}(2014)}]{jon}
\bibinfo{author}{\bibfnamefont{J.}~\bibnamefont{Shlens}},
  \bibinfo{journal}{CoRR} \textbf{\bibinfo{volume}{abs/1404.2000}}
  (\bibinfo{year}{2014}), \urlprefix\url{http://arxiv.org/abs/1404.2000}.

\bibitem[{sup(2019)}]{supplementary}
\bibinfo{journal}{See Supplementary Material for programs and raw data}
  (\bibinfo{year}{2019}).

\bibitem[{\citenamefont{Makhlin}(2002)}]{makhlin}
\bibinfo{author}{\bibfnamefont{Y.}~\bibnamefont{Makhlin}},
  \bibinfo{journal}{Quantum Information Processing}
  \textbf{\bibinfo{volume}{1}}, \bibinfo{pages}{243} (\bibinfo{year}{2002}).

\bibitem[{\citenamefont{Shende} \emph{et~al.}(2004)\citenamefont{Shende,
  Markov, and Bullock}}]{shende}
\bibinfo{author}{\bibfnamefont{V.~V.} \bibnamefont{Shende}},
  \bibinfo{author}{\bibfnamefont{I.~L.} \bibnamefont{Markov}},
  \bibnamefont{and} \bibinfo{author}{\bibfnamefont{S.~S.}
  \bibnamefont{Bullock}}, \bibinfo{journal}{Phys. Rev. A}
  \textbf{\bibinfo{volume}{69}}, \bibinfo{pages}{062321}
  (\bibinfo{year}{2004}),
  \urlprefix\url{https://link.aps.org/doi/10.1103/PhysRevA.69.062321}.

\end{thebibliography}

\end{document}